\documentstyle[aps,twocolumn,epsfig,graphics]{revtex}
\begin{document}
\twocolumn[\hsize\textwidth\columnwidth\hsize
\csname@twocolumnfalse%
\endcsname
\draft
\title{Soliton ratchets}
\author{Mario Salerno$^\dag$  and Niurka R.\ Quintero$^\ddag$}
\address{$^\dag$Dipartimento di Fisica "E.R. Caianiello" and Istituto
         Nazionale di Fisica della Materia (INFM),\\ Universit\'a di
         Salerno, I-84081 Baronissi, Salerno, Italy.\\
         $^\ddag$Grupo de F\'{\i}sica No Lineal (GFNL), Departamento de
         F\'{\i}sica Aplicada I, Facultad de Inform\'atica,\\ Universidad de
         Sevilla, Avenida Reina Mercedes s/n, 41080 Sevilla, Spain.
         }
\date{\today}
\maketitle
\begin{abstract}
The mechanism underlying the soliton ratchet, both in absence and
in presence of noise, is investigated. We show the existence of an
asymmetric internal mode on the soliton profile which couples,
trough the damping in the system, to the soliton translational
mode. Effective soliton transport is achieved when the internal
mode and the external force are phase locked. We use as working
model a generalized double sine-Gordon equation. The phenomenon is
expected to be valid for generic soliton systems. \hskip 1cm PACS
numbers: 05.45.Y, 05.45.-a, 05.60.C, 63.20.Pw
\end{abstract}

]

One of the phenomena which is presently attracting interest both
in physics and in biology is the so called ratchet effect
\cite{Ajd92}. In simple terms, a ratchet system can be described
as a periodically forced brownian particle moving in an asymmetric
potential in presence of damping and periodic driving
\cite{hangggi}. The periodic forcing keeps the system out of
equilibrium so that the thermal energy can assist the conversion
of the ac driver into effective work (direct motion of the
particle) without any conflict with the second law of
thermodynamics. This phenomenon has been found in several physical
\cite{physical} and biological \cite{biological} systems and is
presently considered as a possible mechanism by which biological
motors perform their functions \cite{Ast94}. For ode systems with
damping, additive forcing and noise, the ratchet effect can be
viewed as a phase locking phenomenon between the motion of the
particle in the periodic potential and the external driver
\cite{Barbi}. Ratchet dynamics have been observed also in more
complicated systems such as overdamped $\phi^4$ models
\cite{marchesoni}, chains of coupled particles with degenerate
on-site potentials \cite{savin}, long Josephson junctions with
modulated widths \cite{goldobin}, inhomogeneous parallel Josephson
arrays \cite{trias}, 3D helical models \cite{zolo}, etc. These are
infinite dimensional systems described by continuous or discrete
equations of soliton type, with asymmetric potentials, damping,
and periodic forcing, in which the ratchet phenomenon manifests as
unidirectional motion of the soliton (soliton ratchet). For
overdamped systems one can reduce the soliton ratchet to the usual
single particle ratchet by using a collective coordinate approach
for the center of mass of the soliton \cite{marchesoni}. For
underdamped or moderately damped systems, however, this could be
unappropriate, since the radiation field present in the system can
play an important role for the generation of the phenomenon.

The aim of this Letter is to investigate the mechanism underlying
soliton ratchets both in absence and in the presence of noise. To
this end we use an asymmetric double sine-Gordon equation as a
working model for studying the effect (the phenomena, however,
will not depend on the particular model used). We show that the
asymmetry of the potential induce a spatially asymmetric internal
mode on the soliton profile which can be excited by the periodic
force. In presence of damping, this mode can exchange energy with
the translational mode so that the soliton can have a net motion
under the action of the ac force. In this mechanism, the damping
plays the role of coupling between the internal mode and the
translational mode of the soliton. We find that, for fixed
amplitude and frequency of the ac force, there is an optimal value
of the damping for which the transport (i.e. the velocity achieved
by the soliton) becomes maximal. In this case the frequency of the
internal mode and the one of the external driver, perfectly match
(phase locking). A similar resonant behavior is also observed by
varying the frequency of the ac force, keeping fixed the other
parameters of the system. At very high damping and fixed amplitude
of the forcing, the internal mode oscillation becomes very small,
and the transport due to soliton-ratchet is strongly reduced. At
low damping and higher forcing we find, quite surprisingly, that
current reversals can occur. Finally, we show that soliton
ratchets can survive the presence of noise in the system.

\begin{figure}[tbp]
\centerline{\epsfig{figure=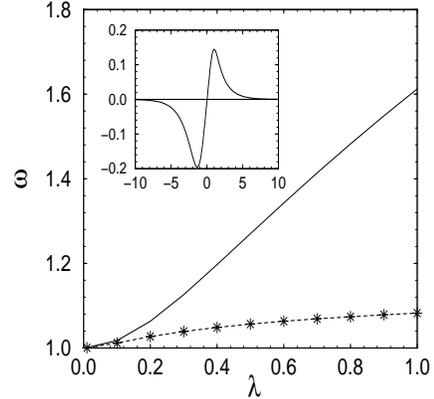,width=2.2in,height=2.2in,angle=270}}
\caption{Small oscillation spectrum versus $\lambda$. Above the
solid line are the phonon's modes. The starts joined by the dashed
line refers to the internal mode. The inset shows the spatial
profile of the internal mode at $\lambda=0.5$.} \label{fig1}
\end{figure}

We start by introducing the following generalized double
sine-Gordon equation
\begin{equation}
\phi _{tt}-\phi _{xx}=- \sin(\phi)- \lambda \sin (2\phi +\theta)
\,  \equiv -\frac{dU(\phi )}{d\phi }, \label{nlkg}
\end{equation}
with the potential $U(\phi )=C-\cos (\phi )-\frac{\lambda }{2}\cos
(2\phi +\theta )$. Here $\lambda $ is the asymmetry parameter,
$\theta $ is a fixed phase and $C$ a constant which fix the zero
of the potential. A discrete version of this equation was
introduced in Ref. \cite{ms85} in terms of a chain of elastically
coupled double pendula assembled by a gear of ratio $1/2$ with a
phase angle $\theta$ between them. For $\lambda =0$ Eq.
(\ref{nlkg}) gives the well known sine-Gordon equation (SGE) with
exact soliton solutions, while for $\lambda \neq 0$ and $\theta
=0$ (mod $\pi $) it reduces to the proper double sine-Gordon
equation (note that in both cases the potential is periodic and
symmetric in $\phi$). In the following we are interested in the
case $\theta \neq 0$ (mod $\pi )$ for which the periodic potential
becomes asymmetric. We shall refer to this case as the asymmetric
double sine-Gordon equation (ADSGE). In particular, we fix $\theta
=\pi/2$ in Eq. (\ref{nlkg}) in order to have maximal asymmetry,
and choose $C=\cos \left(\phi_{0}\right) -{\lambda }/{2}\sin
\left( 2\phi_{0}\right)$,  with $\phi _{0 }=\arcsin
[(1-A)/{4\lambda }]+2n\pi$ and $A=\sqrt{1+8\lambda ^{2}}$, to have
the zero of the potential in correspondence with its absolute
minima  $\phi_{0 }$. We also assume, for simplicity, $\lambda \in
[-1,1]$ to avoid relative minima appearing in the potential.
Besides the mentioned mechanical model, Eq. (\ref{nlkg}) is also
linked to another interesting physical system  i.e. a one
dimensional array of inductively coupled SQUIDs, each consisting
of a loop of a Josephson junction in parallel with a serie of two
identical Josephson junctions. The single element of this array
was studied in Ref. \cite{hangii} in which it was shown that, due
to the ratchet effect, the system can rectify periodic signals. In
analogy with the single particle ratchet, it is reasonable to
introduce in the distributed model,  periodic forcing, damping,
and noise, this leading to the following perturbed ADSGE
\begin{eqnarray}
\phi _{tt}-\phi _{xx} &+&\sin (\phi )+\lambda \cos (2\phi )=  \label{asydsg}
\\
&-&\alpha \phi _{t}+\epsilon \sin (\omega t+\theta_{0})+n(x,t).
\nonumber
\end{eqnarray}
Here $\alpha $ denotes the damping constant, $n(x,t)$ is white
noise with autocorrelation
\begin{equation}
<n(x,t){n}(x^{\prime },t^{\prime })>=D\delta (x-x^{\prime })\delta
(t-t^{\prime }),  \label{noise}
\end{equation}
and  $\epsilon $, $\omega $, $\theta _{0}$ are, respectively, the
amplitude the frequency and the phase of the driver. Travelling
wave solutions of the unperturbed ADSGE, i.e. solutions which
depend on $\xi\equiv {(x-Vt)}/{\sqrt{1-V^{2}}}$, (note that Eq.
(\ref{nlkg}) is Lorentz invariant), can be found by substituting
$\phi \equiv \phi(\xi)$ in  the left hand side of Eq.
(\ref{asydsg}) and equating it to zero. After one integration in
$\xi $ we obtain
\begin{equation}
\int_{\phi_{0}}^{\phi }\frac{d\phi }{\sqrt{2(E+U)}}=\xi -\xi _{0},
\label{integral}
\end{equation}
which gives, after the inversion of the integral at $E=0$
(top of the reversed potential), the $2\pi$-kink (antikink)
solutions as
\begin{equation}
\label{exact} \phi^{\pm}_{K}=\phi_{0}+2\tan^{-1}
\{\frac{sign(\lambda)\,A\,B}{A-1- \eta \sinh \left[\pm
\frac{\xi}{2}\sqrt{{AB}/{|\lambda |}}\right]}\}
\end{equation}
where $\eta=2\lambda\sqrt{2(1+A)}$, and $B=\sqrt{2(4\lambda
^{2}-1+A)}$ (the plus and minus signs refer to the kink and
antikink solutions, respectively). Note that in the limit $\lambda
\rightarrow 0$ Eq. (\ref{exact}) reduces to the well known soliton
solution of SGE. To investigate the existence of internal modes in
the system, we linearize the ADSGE around the solution in Eq.
(\ref{exact}), i.e. we look for solutions of the form $\phi=
\phi^{\pm}_{K}+ \psi$ with
\[
\psi (x,t)=\exp (i\omega t)f(x), \quad f(x) << 1.
\]
This leads to the following eigenvalue problem on the whole line
\begin{eqnarray}
f_{xx}+\left( \omega ^{2}-\cos [\phi^{\pm}_{K}]+2\lambda \sin
(2\phi^{\pm}_{K})\right) f &=&0, \label{eq17}
\end{eqnarray}
with $f_{x}(\pm \infty )=0$, which can be easily solved by
numerical methods for any finite length of the system. In Fig.
\ref{fig1} we report the numerical spectrum of Eq. (\ref{eq17}) as
a function of $\lambda$. We see that except for the sine-Gordon
limit ($\lambda = 0$), there is an internal mode frequency
$\Omega_{i}$ below the spectrum of the phonon band (the zero mode,
existing for all values of $\lambda$, is not plotted for graphical
convenience). In the inset of the figure the shape of the internal
mode is also reported, from which we see that the asymmetry of the
potential induces a spatial asymmetry in $f(x)$. In the presence
of a periodic force this internal mode can be easily excited.
To understand the role of the various elements of the problem,
(i.e. asymmetry of the potential, internal mode, damping, forcing,
and noise) it is better to consider first the zero noise case
(deterministic soliton ratchet). By viewing the soliton as a
string lying on the potential surface $\it{S}(U,\phi,x)$ and
connecting adjacent minima, the following picture of the
phenomenon can be given. If the potential is asymmetric (in $\phi
$) the transition from the top of the potential to one minimum and
from the top of the potential to the other minimum, is also
asymmetric (it will be more rapid for the part of the string lying
on the region where the potential is more stiff). Thus, the
potential asymmetry in $\phi$ induces an asymmetry in space which
can be seen both in the $2\pi$-kink profile and in the internal
mode. In presence of an ac force, but in absence of damping, this
asymmetry will not produce transport, i.e. the string will slide,
without any friction, back and forth on the potential profile
along the $x-$direction. The presence of damping, however,
introduces friction in this sliding, and the part of the string
moving on the stiff part of the potential profile dissipate more
that the other. This asymmetry in the dissipation produces net
motion for the soliton (the string moves in the direction in which
it approaches the potential minimum more smoothly). We can say
that the effect of the damping is to couple the internal mode to
the translational mode. The mechanism underlying deterministic
soliton ratchet can then be described as follows: the ac force
pumps energy in the internal mode which is converted into net dc
motion by the coupling with the zero mode induced by the damping.
>From this picture one can easily predict that in absence of the
internal mode, or in absence of damping, no soliton ratchet can
exist. Moreover, one expects that the maximal effect in transport,
is observed when the internal oscillation and the external force
are synchronized (phase-locked).

\begin{figure}[tbp]
\vspace{1cm}
\centerline{\epsfig{figure=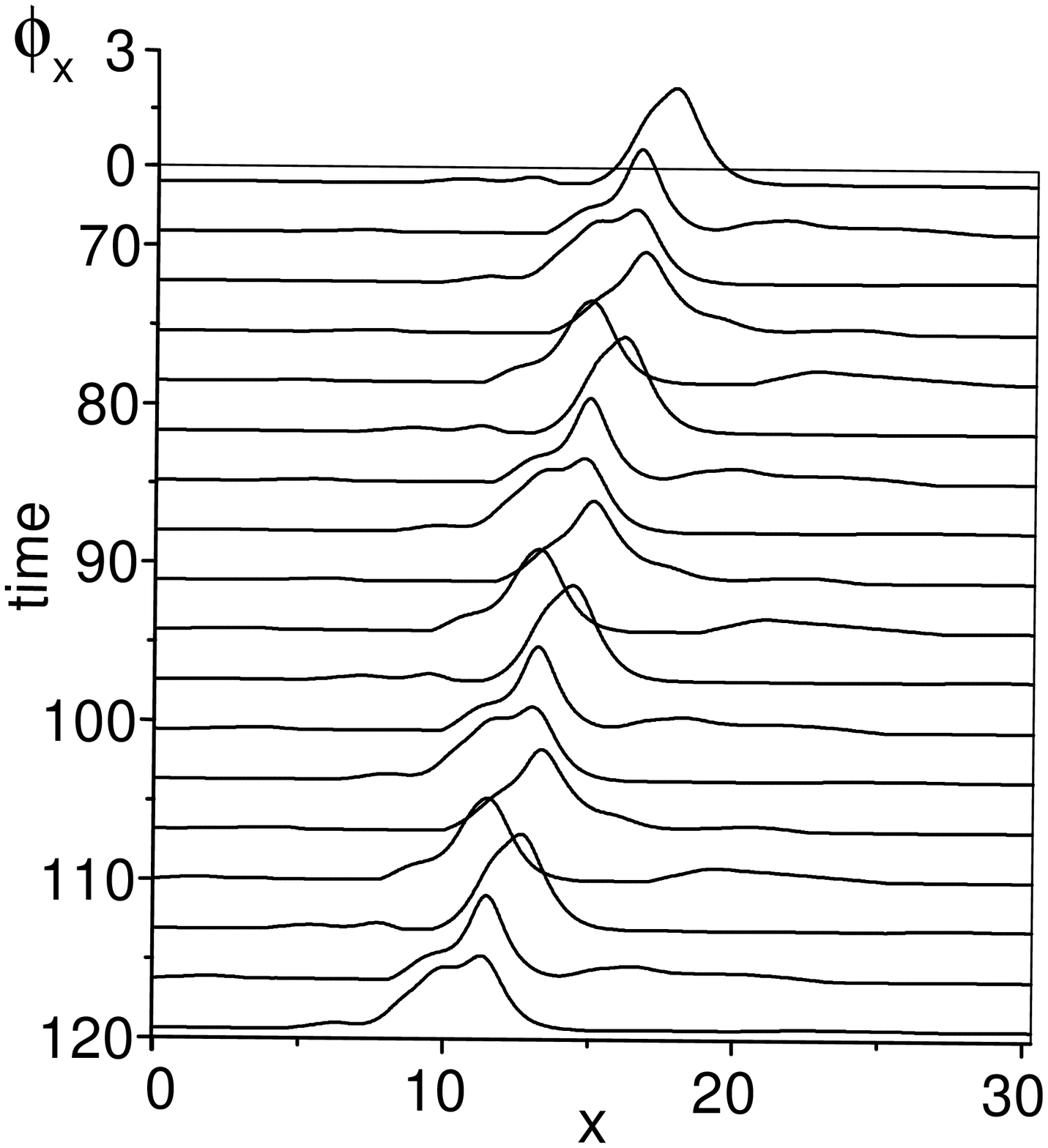,width=2.2in,height=2.2in,angle=0}}
\caption{Time evolution of the x-derivative of the kink profile
while executing ratchet dynamics. The parameter values are
$\lambda=0.5$, $\theta_{0}=\pi/2$, $\alpha=0.6$, $\epsilon=0.8$,
$\omega=0.8$, and $D=0$.} \label{fig2}
\end{figure}

\begin{figure}[tbp]
\centerline{\epsfig{figure=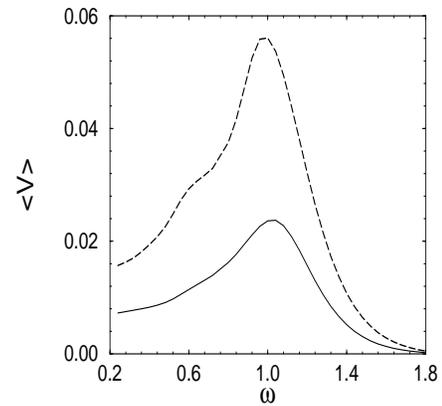,width=2.2in,height=2.2in,angle=270}}
\caption{Mean velocity of the kink center of mass versus the
frequency of the ac force for $\epsilon=0.4$ (solid line) and
$\epsilon=0.6$ (dashed line). The other parameter are fixed as in
Fig.(\ref{fig2}) except for the asymmetry parameter which is
$\lambda=-0.5$.} \label{fig3}
\end{figure}

\begin{figure}[tbp]
\centerline{\epsfig{figure=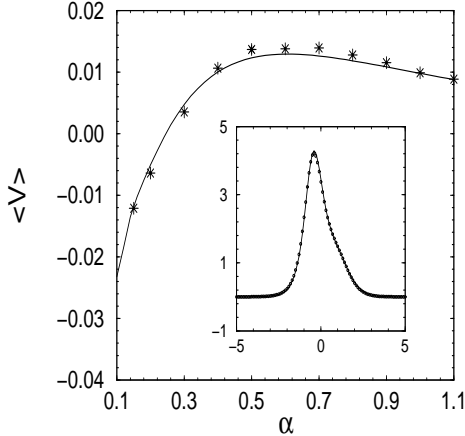,width=2.4in,height=2.4in,angle=270}}
\caption{Mean velocity of the center of mass of the kink versus
$\alpha$, for parameter values $\lambda=-0.5$, $\theta_{0}=\pi/2$,
$\omega=0.4$, $\epsilon=0.5$. The continuous curve refers to the
case $D=0$ while the stars refer to the nondeterministic case
$D=0.01$. The continuous and dotted curves in the inset refer to
two soliton $\phi_x$ profiles at $D=0$, taken in the co-moving
frame and separated in time by one period of the driver.}
\label{fig4}
\end{figure}

In order to confirm this picture, we have performed direct
numerical integrations of the ADSGE for different values of the
system parameters. First, we have checked that in the SGE limit,
i.e. when $\lambda=0$, the ratchet dynamics does not exists. This
agrees with the fact that in this case there is no asymmetry in
the potential and no internal mode. At this point we remark that
the dynamics of a SG kink subject to a periodic force was also
investigated in Ref. \cite{olsa}, in which it was shown that in
absence of damping the kink can acquire a finite velocity
depending on the initial phase of the ac force. Net motion of a
SGE  soliton was shown to be possible also in presence of a small
damping, if the ac force excites a phonon mode which exchange
energy with the soliton \cite{rc}. These cases, however, should
not be confused with soliton ratchets since they strongly depend
on initial conditions (if one average on initial conditions the
transport disappears). Moreover, in contrast with soliton
ratchets, these effects exist only at zero or at very low damping.
Second, we have checked that for $\lambda\neq 0$ (asymmetric
potential) but in absence of damping, soliton ratchets also do not
exist (for brevity we will not expand on these cases here). From
this analysis we conclude that, in analogy with the deterministic
single particle ratchets, the asymmetry of the potential, the
damping and, obviously the periodic forcing, are crucial
ingredient for soliton ratchets. In Fig. \ref{fig2} we show a
prospectic view of the soliton ratchet dynamics as obtained from
numerical integration of Eq. (\ref{asydsg}). We remark that the
direction of the motion is fixed by the asymmetry of the potential
and can be inverted by changing the sign of $\lambda$. We see
that, except for the soliton profile which is wobbling, no phonons
are present in the system. This confirms the relevance of the
internal mode in the phenomenon. We also find that for some
parameter values, the net motion is more effective. To investigate
the dependence of the phenomenon on parameters, we have performed
numerical simulations of Eq. (\ref{asydsg}) both by fixing all
parameters and changing $\omega$, and by fixing all parameters and
changing $\alpha$. In Fig.\ref{fig3} we show the average velocity
(computed by using an integration time $t=1000$) of the soliton
center of mass versus the frequency of the external driver, for
two different values of the amplitude of the ac force (the low
damping part of the curves was not computed due to the longer
integration times required in this case). From this figure we see
that $\langle V \rangle$ has a maximum at $\omega \sim 1$, (i.e
$\omega=1.04$ and $\omega=1$ for $\epsilon=0.4, \epsilon=0.6$,
respectively), this being very close to the internal mode
frequency $\Omega_{i} \approx 1.0562$ (the discrepancy is within
the numerical accuracy of our numerical scheme). By increasing the
amplitude of the forcing, the dynamics gets more complicated
(breather-like excitations can appear) and the resonance peak in
frequency more pronounced.
A similar resonant behavior is expected to exists also as function
of the damping. When the damping is very high, indeed, the
internal mode is almost suppressed by the damping, while when it
is very low the coupling between the internal mode and the
translational mode is very small, both cases giving minimal
transport. In between these extremes, a value of the damping which
allow the internal mode to synchronize with the external driver
and optimize transport, should then exist. This is what we observe
in Fig. \ref{fig4} where the velocity vs damping is reported as a
continuous curve. Although the ratchet velocity can be slightly
increased by optimizing parameters, it remains small in comparison
to the velocities achieved under the action of dc forces. This can
be due to the fact that the internal and the translational mode
seems to couple only to second order in a perturbation scheme. To
increase the velocity one could increase the amplitude of the ac
force, but this is limited by the threshold above which the system
becomes chaotic. In practical applications, however, a small drift
velocity is sufficient to manipulate solitons under the action of
external ac forces.
To show that the internal mode is phase locked with of external
driver we have plotted in the inset of this figure the $\phi_x$
profiles in the co-moving frame (i.e. the drift motion was
subtracted) at two fixed times $t_{1}=211.6$ (solid line) and
$t_{2}=227.3$ (circles) separated by one period
$T=2\pi/\omega=15.7$ of the driver. We see that the profiles
overlap each other, i.e. the oscillation on the kink profile is
perfectly synchronized with the external driver (phase locking).
>From Fig. \ref{fig4} we see, quite surprisingly, that at low
damping current reversals can occur (note that the average
velocity becomes negative for $\alpha$ less than $\alpha_{cr}\sim
0.24$). We find that the value of $\alpha_{cr}$ increases as the
amplitude of the driver is increased. The occurrence of this
phenomenon, which resemble the one observed in single particle
ratchets at low damping \cite{Barbi,cha95}, seems to be related
more to the phonon-soliton interaction than to the internal mode
mechanism described above (at low dampings a complicate
transferral of energy between phonons, internal mode, and
translational mode, can arise). To understand this phenomenon,
however, a more detailed study is required.

Finally, we have investigated the effect of the noise on
deterministic soliton ratchets. A preliminary analysis shows that
for low noise amplitudes the effect of the noise on the phenomenon
is minimal, in the sense that the dynamics gets dressed by the
noise, but after averaging on the noise, almost the same soliton
mean velocity $\langle V\rangle$ is obtained. This is shown by the
stars  in Fig. \ref{fig4} which represent the numerical values of
$\langle V\rangle$ calculated in presence of a noise of amplitude
$D=0.01$. The fact that soliton ratchets can survive the presence
of weak amplitude noise can be understood as consequence of the
structural stability of phase locking phenomena against "small"
fluctuations. This indicates that the phenomenon can exist also in
real systems  such as one dimensional arrays of inductively
coupled SQUIDs. The mechanism of soliton ratchets discussed in
this Letter is expected to be valid also for other soliton
systems.

\vskip .2cm We thank M.Barbi and M.R. Samuelsen for interesting
discussions. Financial support from the European grant LOCNET n.o
HPRN-CT-1999-00163 is acknowledged.


\begin{references}

\bibitem{Ajd92}A.~Ajdari and J.~Prost, {\em C. R. Acad. Sci. Paris}
Ser.2, 315 (1992) 1635; M.~O. Magnasco, Phys. Rev. Lett., {\bf 71}
(1993)1477; C.~R. Doering, Il Nuovo Cimento, {\bf 17} (1995) 685.

\bibitem{hangggi}P. H\"anggi and R. Bartussek, in {\it Lecture notes in
Physics}, Ed.s J. Parisi et al., Springer, Berlin, {\bf 476},
(1996); see also the review of P. Reimann, in cond-mat/0010327.

\bibitem{physical}  L.~Gorre, E.~Ioannidis, and P.~Silberzan,
{\em Europhys. Lett.}, {\bf 33} (1996) 267; L.~Gorre-Talini,
S.~Jeanjean, and P.~Silberzan, {\em Phys. Rev. E}, {\bf 56} (1997)
2025; O.~Sandre {\em et al}, Phys. Rev. E, {\bf 60} (1999) 2964.

\bibitem{biological} J.~S.~Bader {\em et al.},
{\em Proc. Natl. Acad. Sci. USA}, {\bf 96}(1999) 13165; N.~J.
Cordova, B.~Ermentrout, and G.~F. Oster, {\em Proc. Natl. Acad.
Sci. U.S.A.}, 89(1):39, 1992; F.~J{\"{u}}licher, A.~Ajdari, and
J.~Prost, Rev. Mod. Phys. {\bf 69} (1997) 1269.

\bibitem{Ast94}R.~D.~Astumian and M.~Bier, Phys. Rev. Lett. {\bf 72} (1994) 1766.

\bibitem{Barbi}  M. Barbi and M. Salerno, Phys. Rev. E, {\bf 62} (2000) 1988.

\bibitem{marchesoni}F. Marchesoni, Phys. Rev. Lett. {\bf 77}, (1996) 2364.

\bibitem{savin}A.V. Savin, G.P. Tsironis, A.V. Zolotaryuk, Phys.Lett. A {\bf 229} (1997)
279; Phys. Rev. E {\bf 56} (1997) 2457.

\bibitem{goldobin}  E. Goldobin, A. Sterck, and D. Koelle,
Phys. Rev. E {\bf 63} (2001) 031111.

\bibitem{trias}E. Tr\'{\i}as, J.J. Mazo, F. Falo, and T.P.
Orlando, Phys. Rev. E {\bf 61} (2000) 2257.

\bibitem{zolo}A.V. Zolotaryuk, P.L. Christiansen,
B.Nord\'en, and  A.V. Savin, Cond. Mat. Phys., Vol.2, No 2(18),
(1999) 293.

\bibitem{ms85}Mario Salerno, Physica\ D {\bf 17}, (1985) 227.

\bibitem{hangii}I. Zapata, R. Bartussek, F. Sols, and P. H\"anggi,
Phys. Rev. Lett. {\bf 77}, (1996) 2292.

\bibitem{olsa}  O.\ H.\ Olsen and M.\ R.\ Samuelsen, Phys.\ Rev.\ B {\bf 28} (1983) 210.

\bibitem{rc}  N.\ R.\ Quintero, A.\ S\'{a}nchez, and F.\ G.\ Mertens
Phys.\ Rev.\ E {\bf 62}, Rapid Comm. (2000) 1.

\bibitem{cha95}  J.F. Chauwin, A.~Ajdari, and J.~Prost,
Europhys. Lett. {\bf 32} (1995) 373; J.L. Mateos, Phys. Rev. Lett.
{\bf 84} (2000) 258.


\end{references}
\end{document}